# Improving Rehabilitative Assessment with Statistical and Shape Preserving Surrogate Data and Singular Spectrum Analysis

T. K. M. Lee
School of Electrical & Electronic Engineering
Singapore Polytechnic, Singapore
School of Information Technology
Monash University Malaysia
tlee@sp.edu.sg

H. W. Chan
Dept of Aerospace Engineering
University of Glasgow
Singapore
2427293C@student.gla.ac.uk

K. H. Leo
Singhealth Office for
Organizational Transformation
Singapore
leo.kee.hao@singhealth.com.sg

E. Chew
Division of Neurology
National University Hospital
Singapore
effie_chew@nuhs.edu.sg

Ling Zhao
Division of Neurology
National University Hospital
Singapore
ling_zhao@nuhs.edu.sg

S. Sanei
School of Science & Technology
Nottingham Trent University
United Kingdom
saeid.sanei@ntu.ac.uk

*Abstract*—Time series data are collected in temporal order and are widely used to train systems for prediction, modeling and classification to name a few. These systems require large amounts of data to improve generalization and prevent over-fitting. However there is a comparative lack of time series data due to operational constraints. This situation is alleviated by synthesizing data which have a suitable spread of features yet retain the distinctive features of the original data. These would be its basic statistical properties and overall shape which are important for short time series such as in rehabilitative applications or in quickly changing portions of lengthy data. In our earlier work synthesized surrogate time series were used to augment rehabilitative data. This gave good results in classification but the resulting waveforms did not preserve the original signal shape. To remedy this, we use singular spectrum analysis (SSA) to separate a signal into trends and cycles — to describe the shape of the signal — and low level components. In a novel way we subject the low level component to randomizing processes then recombine this with the original trend and cycle components to form a synthetic time series. We compare our approach with other methods, using statistical and shape measures and demonstrate its effectiveness in classification.

I. INTRODUCTION

The feats of deep learning as part of artificial intelligence in the areas of prediction, simulation and classification have been exceptional. This has been achieved by training neural networks (NN) on high dimensional data using large amounts of data to achieve robustness, improve generalization and prevent overfitting of results. Even then the training has employed some form of data augmentation.

Much of these advances have been made using images, as huge caches of such data have been shared online. But this is not always the case for other fields, especially for time series where data collection is limited due to constraints in privacy, expense, time, physical access and so on.

Thus there is a need to augment time series data to effectively train machine learning systems. Besides, other uses for synthetic time series are for testing null-hypotheses, validating new methodologies and producing ensembles for event attribution. Recently there have been a large number of works on synthesizing time series which employ generative methods using various types of generative adversarial neural networks (GAN) as reviewed in [8]. These are typically resource intensive both in terms of training and deployment. Traditional methods of time series synthesis take up less resources, their mechanisms more easily understood also in terms of training and deployment. This would be useful for example, for resource constrained embedded systems. These considerations motivate for our approach.

Synthesizing a time series requires an understanding of the original data generating process. Thus, for processes in the steady state, it is straightforward to use the mean and standard deviation to synthesize the data as the temporal property of data is not of the main concern. In many cases adding a small amount of random noise is useful where noise is present in the process for example from sensors and biomedical applications. However in long time series important transient events occur during the start of a process or prior to exceptional events like process failure. Together with many other naturally occurring short time series, these have been successfully modelled by linear stationary stochastic processes.

From these considerations we see that synthetic time series should exhibit i) fidelity — preserving the basic statistical properties of the original data ii) diversity — have a useful spread of characteristics so the systems can generalize better about the data: for this we use the *shape* of the time series. Also, we will use the notions of fidelity and diversity in an empirical way.

Methods to synthesize data may not preserve certain original time series features but as we mention later, they have been successfully used to train machine learning systems. It would be useful to explore how to improve on retaining these

This version of the paper under the same title, acknowledges the data source and the funding for current research using this data.

features. We focus on accelerometric data obtained from a rehabilitation setting

In Section II we describe the motivation for our approach as well as the background material. Section III outlines our physical setup followed by the theory and approach in Section IV. Our experimental results are presented in Section V and we summarize our discussion with conclusions in Section VI. For the rest of the paper we use the terms time series and signal interchangeably.

## II. BACKGROUND WORK

In this section we review the data augmentation approaches, ways to assess their fidelity and diversity and evaluating their effectiveness in classification.

In our earlier work we used surrogate data to augment training data [1] to successfully train a NN based classifier. However the waveforms of the surrogate time series in many cases were quite unlike the shape of the original signal as seen in Fig 4E. As mentioned in Section I, there are situations where the synthesized data should follow the shape of the original. This is the motivation to improve on the surrogate data approach; the preservation of the original signal shape. The concept of the shape of a signal is empirical but some ideas that spring to mind are those of the general trend of the data as seen in Fig 4A.

With this in mind, we note that a time series may be composed of trend, cycle, seasonal and irregular components [2]. We can do no worse than taking the trend, cycle and seasonal parts of a time series as its *shape*.

In this respect Singular Spectrum Analysis (SSA) extracts trend, cycle and noise (or irregular) components of a time series in a non-parametric way. SSA has been used extensively in biomedical applications[3][4][5] and work by Vautard et al. [6] focuses on short, noisy signals which are applicable to our data.

To create diversity in synthetic time series, the method of surrogate data uses bootstrapping, resampling from the original data with replacement. One of the key motivations for surrogate data is for the synthesized data to preserve the basic statistical features of the original namely: its mean, standard deviation and power spectrum or autocorrelation. The work by Theiler et. al [5] has spawned large number of variations as described in [7]. As mentioned earlier, it would be simple to just shuffle the time series data randomly taking note of the mean and variance of the original data. But to preserve the power spectrum requires a more nuanced approach which Theiler et al. [8] implemented as the Amplitude Adjusted Fourier Transform (AAFT).

We consider to retain the *shape* of a time series as just defined and create diversity by randomizing its irregular component.

The closest use corresponding to our method is by Kostenko and Vasylyshyn who explored how to improve the effectiveness of spectral signal analysis and in [9] they describe how to generate surrogate data based on the noise component of a signal decomposed by SSA.

An interesting variation on this method is to replace the irregular data component with actual signals from another domain rather than noise [10].

In the review of time series augmentation by Iwana and Uchida [11] they reckon that techniques for augmenting image data are not easily transferable to time series. For example, rotating or flipping data could cause severe distortion in the temporal sequence of data. However, certain other methods could be useful. They provide a taxonomy of techniques that covers four broad areas:

i) Random variations in signal amplitude, timings and frequencies.
ii) Mixing of signal patterns by combining amplitude, time or frequency from the existing signals linearly or otherwise.
iii) Generative methods using statistical properties or working in manifold space such as those using GANs.
iv) Decomposition of the data into other components (besides frequency) like seasonal, trend and residue, where only the residue is randomized. This is then recombined with the other components to produce the synthetic data.

They have also prioritized some techniques which work well when tested with a variety of classifiers especially those applying Convolutional Neural Networks (CNN) such as what we are using. We include two of these methods as a comparison, which are window slicing and warping [12].

Since only SSA gives us a basis for what is a shape, we call upon another measure of signal fidelity which is its *similarity* to the original data. A widely used measure is Dynamic Time Warping (DTW) [13]. The DTW algorithm non-linearly warps the data in time to match two time series in an optimal way. The prominent output of this process is a distance which measures how close two time series are, being smaller for two time series that are similar to each other in shape.

Classification
Using neural networks to classify time series have proven very successful, starting with early approaches such as transforming 1D data to 2D images as the main advances in machine learning was involved images. In our knowledge, we were the first to effectively use surrogate data to train systems to classify time series [1]. We used transfer learning from large 2D CNNs with millions of parameters. Subsequently we showed that a 1D CNN and a Long Short-Term Memory (LSTM) neural network were able to give very good results with the amount of data being augmented 50 fold [14]. However, 1D CNNs have proven successful for this task as well with fewer computing resources.

In 2D CNNs, the 2D filter coefficients that have been effective for classifying images are determined as part of the training process. In the same way 1D CNNs learn the best 1D filters for classification. These filters, also called feature maps can be readily visualized, giving a good idea of the important features that aid classification. The LSTM networks can give better results but its structure is more complex and training requires more resources.

## III. DATA COLLECTION AND SETUP

Our dataset comprises data from an accelerometer embedded in a 10 cm cube. The cube is moved by subjects in a manner prescribed by a rehabilitative test. Three channels of data from the triaxial device are digitized to an 8 bit resolution at a sampling rate of 30 Hz — with more details in[15]. The movements are visually scored by clinical staff and starts from a score of 3 for the best quality of movement to 1 and a score of 0 is the inability to complete the test. In Fig. 1 we see how the cube is gripped, held vertically and moved. The movement is highly constrained and can be viewed as a one-dimensional movement in 3D space.

The data were recorded from 34 patients who have had a history of stroke and undergone rehabilitation. The medical trial was conducted in a hospital over two months.

From these subjects, 78 sets of data were recorded. Of these 31 scored at 3, 38 scored 2, 6 scored 1 and 3 scored 0.

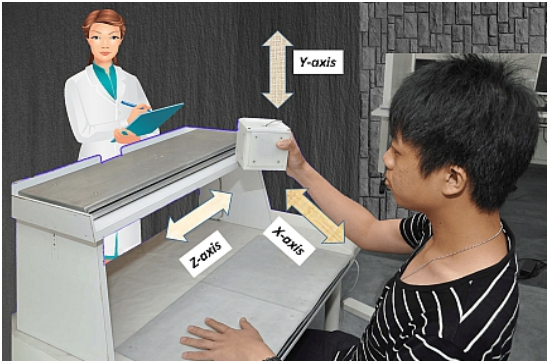

**Fig. 1.** Rehabilitative test with scorer in background.

Thus we consider this a classification exercise where the scores are subjectively given by the scorer. This dataset is challenging due to the lack of visually discernible inter and intra class features as well as visual subjective scoring over time.

*A. Dataset augmentation*

To reduce statistical bias we generate synthetic data so that each class has roughly the same number of time series. We denote various versions of the synthetic dataset by the fold increase of time series being synthesized - so a 10-fold increase would have 10 time series synthesized from an original as shown in Table 1. Data with scores of 0 and 1 are augmented more to equalize the distribution of data.

*B. Data preprocessing*

The motion data captured over a session are manually segmented and data outliers due to noise compensated for.

**Table 1** Examples of data set augmentation. Quantities in the center cells are the number of time series: x-fold is the fold increase. T.S. refers to time series. For example for a score of 3, there were originally 31 time series. In the 10-fold dataset each of these time series was augmented 10 times so there are 310 time series with score of 3.

| Score | Original - # T. S. | 10-fold - # T. S. | 50 - # T. S. |
|---|---|---|---|
| 3 | 31 | 310 | 1550 |
| 2 | 38 | 380 | 1900 |
| 1 | 6 | 360 | 1800 |
| 0 | 3 | 360 | 1800 |
| Total samples | 78 | 1410 | 7550 |

## IV. THEORY AND APPROACH

The theory for the components of our system is covered here with SSA then surrogate data. We briefly describe the CNN used and our classification approach.

*A. Singular Spectrum Analysis*

SSA is a subspace analysis method originally developed for single channel time series analysis. In this section we describe Basic SSA following the work by Vautard et al. [3] where for a time series at time $t$ the data is represented by a vector $x(t) = \{x(t): t=1...N\}$ with $N$ sequential, equally spaced time intervals. A sliding window of length $M < N$ embeds this series into a trajectory matrix $\mathbf{Y}$, of size $(N - M + 1) \times M$ where its first and second rows are vectors:

$x_1 = [x(1), x(2),..., x(M)]$
$x_2 = [x(2), x(3),..., x_i(M+1)]$ and for row $N-M+1$
$x_{N-M+1} = [x(N-M+1), x(N-M+2),..., x(N)]$

By concatenating the so called lagged vectors, the trajectory matrix is:

$$\mathbf{Y} = \begin{bmatrix} x_1 \\ x_2 \\ ... \\ x_{N-M+1} \end{bmatrix}$$

The Singular Value Decomposition (SVD) of $\mathbf{Y}$ starts by forming the $M \times M$ covariance matrix $\mathbf{C}$:

$$\mathbf{C} = \mathbf{Y}^T\mathbf{Y} / N$$

where $^T$ is the transpose operator. The diagonalization of $\mathbf{C}$ produces sorted scalar eigenvalues $\lambda$ and eigenvectors $e^k$ (length $M$). The singular values of $\mathbf{Y}$ are then $\sqrt{\lambda}$ and these are used to form principal components ($PC$), the $k^{th}$ $PC$ being a vector of length $N - M + 1$, is given by:

$$a_t^k = \sum_{j=1}^{M} x(t+j) e_j^k \qquad \text{for } 0 \leq t \leq N - M \qquad (1)$$

To view the effects of the decomposition a reconstructed signal component ($RC$) can be generated. This vector of length $N$ corresponding to the $k^{th}$ eigenvalue has its components constructed so at sample instance $t$ we have:

$$RC(t) = \frac{1}{M} \sum_{j=1}^{M} \sum_{k \in K} a_{t-j}^k e_j^k \qquad \text{for } M \leq t \leq N - M + 1$$

$$= \frac{1}{t} \sum_{j=1}^{t} \sum_{k \in K} a_{t-j}^k e_j^k \qquad \text{for } 1 \leq t \leq M - 1$$

$$= \frac{1}{N-t+1} \sum_{j=t-N+M}^{M} \sum_{k \in K} a_{t-j}^k e_j^k \qquad \text{for } N - M + 2 \leq t \leq N$$

where $K$ is the set of $PC$s used for reconstruction. The different equations are needed to cater for the beginning and end conditions of the embedding operation. We use a value of $M = 17$ as it has proven successful in our earlier work [11].

*B. Significant eigenvalues*

The SVD returns a set of $M$ eigenvalues which are sorted to show the contribution of that eigenvalue to the variation of the data. For convenience we use the term λ rather than √λ and in plotting their values we obtain a scree plot, similar to the rubble at the foot of a mountain as seen in Fig. 2. The slope of the plot changes quickly in the first few λs, and settles to a gentle slope for the rest. This identifies the significant λs and eigenvectors which are attributable to trend and seasonal data. The remaining values are assigned to the irregular part of the time series, which in many instances is treated as noise. In this paper we obtain the significant eigenvalues based on Relevant Dimension Estimation (RDE) by Braun et al. [16].

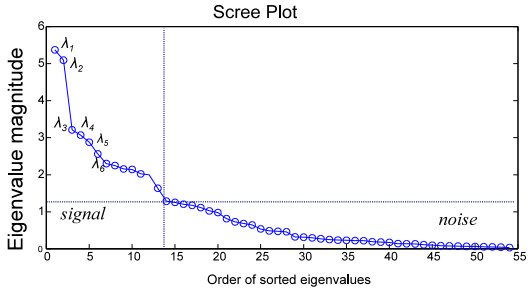

**Fig. 2.** Scree plot - first six eigenvalues indicated

### C. Randomizing with Surrogate Data and Windows Slicing / Warping

In this section we describe the algorithm to generate surrogate data then briefly, the window distortion methods. Previously we input the entire signal but here we only input the *irregular* component of the signal. In the following description, the original time series is a vector *x* of length *N* and all data sorts are in ascending order. All letters in bold italics are vectors.

1. Sort *x* to $x_s$ with the indexes of the sorted values $i_x$
   Sort $i_x$ to form the rank index vector $i_{rx}$.
2. Generate a vector of length *N* from a normal distribution.
   Sort this vector to give *rv*.
3. Permute *rv* using $i_{rx}$ as indexes to create a new vector *rrv*.
4. Compute the Fourier transform of *rrv* as *ft*.
   Generate a vector *φ* of length *N*/2 of random angles from the range [0 2π].
   Compute the phase randomized vector $ft_r$:
   For the first half of $ft_r$ multiply the first half of *ft* by exp(*iφ*)
   For the second half of $ft_r$, take the flipped complex conjugate of the first half.
5. Take the inverse Fourier transform of $ft_r$ to form vector *s*.
6. Sort *s* to obtain indexes $i_s$ ; sort $i_s$ to form rank indexes $i_{rs}$
7. Permute $x_s$ using $i_{rs}$ as indexes to get the surrogate vector.

Window slicing retains 90% of data at random starting points of the signal while warping expands or contracts a random window of 10% of the data in the time series. In both cases the data samples were interpolated to retain the original length of the time series.

### D. 1D Convolutional Neural Network

Generally, the outputs $y_i$ of a NN can be represented by:

$$[g_o(\sum_m w^L_{nm} [\cdots [g_2(\sum_j w^2_{kj}[g_1(\sum_i w^1_{ji} x_i + b^1_j)] + b^2_k]\cdots]_m + b^L_o)]_o$$

where $x_i$ are the inputs, *b* are the bias weights, *w* the interconnection weights and *g* the activation functions in the layers and $g_o$ denotes the output layer. The superscripts denote the layer and subscripts the neuron and connections between them. The selection of the inputs, weights, activation functions, range of summations will vary according to the architecture of the NN. For example in the CNN, the first few convolutional layers have no bias and the weights are filter coefficients. Not all inputs are connected to all the activation functions which just sum the weighted inputs. The subsequent pooling layers have fixed weights of one and the activation functions perform summarizing functions like averaging. The final layers would have ReLU activations and fully connected NN layers.

As in most NN architectures the input needs to be a fixed length vector. We have fixed this to 91, the median of our variable length data. The time series is zero-padded or truncated to this length before being used to train the 1D CNN.

Here it is only required to train on 3 classes. We configured a 1D CNN with 3 sets of convolutional and pooling layers, then 2 sets of dense and dropout layers, all with ReLU activations. The final dense layer used softmax activation with 3 outputs resulting in a network of 26,800 parameters. We used a mini-batch size of 20 over 20 epochs with the ADAM adaptive weight update rule: this starts from a learning rate of 0.001, decaying at a rate of 1E-6 and uses accuracy as the learning metric. In the augmented dataset, 80% was used for training, 20% for validation and the original 78 time series used for testing.

### E. Methodology

We denote our proposed system as the SSA/Surrogate method. The input signal is decomposed by SSA into trends, cycles and low level components. The trend-cycle and seasonal data or shape of the signal is passed on unchanged and thus preserved. The low level signal is subjected to surrogate data processing. Both of these signal components are recombined to form the synthetic time series as shown in Fig. 3.

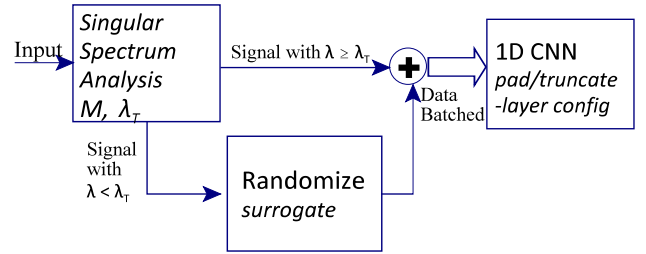

**Fig. 3.** Block diagram of the data augmentation data with parameters in italics described in the text. Input decomposed using SSA (left) into general shape and low level component - which are subject o randomizing processes (middle). Data is recombined, batched and used to train the classifier (right).

This process is repeated for the number of specified fold increases. This process was also applied for the window slicing and warping methods which did not have any signal decomposition.

## V. RESULTS

In this section we compare the statistical and shape preserving abilities of our algorithms followed by their effectiveness in classification. It should be noted that the statistical values and signal shapes are only indicative and will differ slightly for each iteration of the generating algorithm.

We use the changes in four numerical measures to compare the preservation of features in the synthesized signals. Three are statistical.

i) Signal mean: The difference between the simple average of the original and synthesized signal values as a percent of the average of the original signal.

ii) Signal standard deviation: Same as i) except we compute the standard deviation instead.

iii) Signal autocorrelation function (ACF): This is the root mean squared difference between the ACF of the two signals.

The fourth measure gauges the similarity between the two signals using:

iv) DTW: The signals are normalized to zero mean and unity standard deviation. After computing the DTW this is further normalized by dividing by the length of the signal.

A fifth comparison is empirical and done by showing plots of the synthetic time series overlaid on the original signal from the movement data from subject *p14* performing the first trial of the test. Fig. 4A shows the original signal and the SSA decomposition into trend-cycle and irregular components.

TABLE III DIFFERENCES OF STATISTICS AND SIMILARITY BETWEEN ORIGINAL AND SYNTHESIZED TIME SERIES

| Difference | SSA/Surr | Window Scaling | Window Warping | Surrogate only |
|---|---|---|---|---|
| Δ mean % | 0 | -0.7 | -18.24 | 0 |
| Δ std % | 0.52 | -2.74 | -3.89 | 0 |
| Δ autocorr | 0 | 5 | 5 | 0.3 |
| DTW% | 20.19 | 11.63 | 17.39 | 34.72 |

In general we see that SSA/Surrogate best preserves the statistical properties as it produces the smallest changes in values. While the DTW (distance) is larger than the window distortion methods, it is an indication that it provides good diversity but still follows the original signal shape in Fig. 4B. In comparison the window distortion methods does not preserve the statistical data as well although they have good similarity measures.

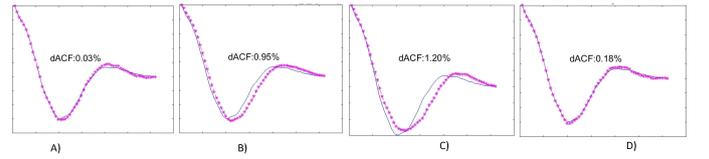

**Fig. 5.** ACF plots - original signal in solid blue, synthesized in dotted lines. A) Windows Scaling B) Windows Warping C) SSA/Surrogate D) Surrogate only

Finally we take the synthesized data and perform classification. In Table IV we see our SSA/Surrogate method gives the best results, being able to predict the condition of a subject with 100% accuracy with just a 25 fold increase of the time series data.

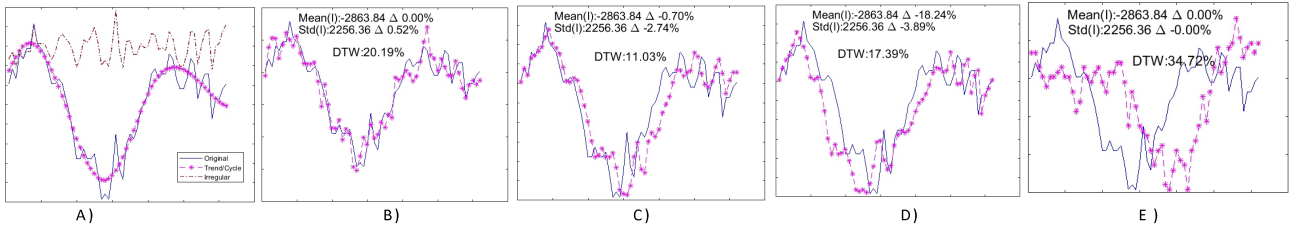

**Fig. 4.** Examples of waveforms: all solid blue lines are original signal. Overlays are A) dotted: SSA derived trend/cycle - dashed: irregular B) dotted: SSA/surrogate C) dotted: windows scale D) dotted: windows warp E) dotted: surrogate only.

As defined, the *shape* of the original signal (dotted line) is consistent with the gross movement of the accelerometer in the cube in Fig. 1. There are some higher frequency components which are a combination of tremors as well as electronic noise [12] and as mentioned earlier we term this as the irregular signal.

Fig. 4B shows the waveform from our work and together with panels C) and D) show a general adherence to the original shape. In comparison panel E) shows the non-shape preserving surrogate data from our previous work [11]. We next show the ACF plots in Fig. 5 and the summary comparison information in Table III. This table shows the values of the percent differences in mean, standard deviation, ACF and DTW between the original and synthesized signal.

The window distortion methods also performed well and we confirm that the surrogate only method performed well in classification even with low shape conformance,.

## VI. CONCLUSION

In summary we attempted to classify the condition of a subject based on their performance in a rehabilitation test using data from a triaxial accelerometer. We recall that from Section III that we are attempting to objectify what is an essentially a subjective score. This score is awarded by different scorers over time and not normalized.

TABLE IV ACCURACIES OF FOLD INCREASE WITH FOUR TIME SERIES SYNTHESIS METHODS

| Synthesis method | SSA/Surrogate | | | | Window Scaling | | | | Window Warp | | | | Surrogate | | | |
|---|---|---|---|---|---|---|---|---|---|---|---|---|---|---|---|---|
| Fold increase | 50x | 25x | 10x | 5x | 50x | 25x | 10x | 5x | 50x | 25x | 10x | 5x | 50x | 25x | 10x | 5x |
| Predict accuracy | 1 | 1 | 0.933 | 0.892 | 0.987 | 0.987 | 0.933 | 0.933 | 0.96 | 0.853 | 0.933 | 0.973 | 0.893 | 0.822 | 0.776 | 0.781 |
| Validate accuracy | 0.998 | .994 | 0.997 | 0.952 | 0.99 | 0.994 | 0.971 | 0.914 | 0.981 | 0.876 | 0.97 | 0.943 | 0.932 | 0.912 | 0.901 | 0.922 |
| Train accuracy | 1 | 1 | 1 | 0.923 | 1 | 0.999 | 1 | 1 | 0.979 | 0.912 | 0.951 | 0.948 | 0.941 | 0.932 | 0.91 | 0.881 |

By augmenting our data in a shape preserving way, we achieve excellent results in training a 1D CNN to accurately score the movement of subject. We also show that using our approach, much fewer parameters are needed to train a smaller network. Using surrogate data only, our earlier 1D to 2D approach [1] used millions of parameters with 100 fold augmentation to achieve accuracies of around 97%. Another early approach [11] used a 1D CNN with 140,735 parameters and 50 fold augmentation to achieve this.

A useful outcome in our method is that it produces a synthetic time series that preserves important signal characteristics viz.: the mean, variance, autocorrelation, and shape which are important for physical modeling and simulation purposes. In our case it would be the movement of picking up and lowering of an object.

Although our method has been used with rehabilitative data, the results are very encouraging as we have compared it favorably with other popular time series augmentation methods. Thus it can be used for other types of accelerometric data. Further work would involve testing other hyper parameters and randomizing schemes to generate more usable synthetic data.

ACKNOWLEDGMENT

We thank Brian Iwana, Eric Breitenberger, Prof D. Kugumtzis and his team for making their software available. We also thank the anonymous reviewers for their comments which have improved the quality of this paper. The dataset used in this study was funded by Ministry of Education of Singapore grant 2010MOE-IF-005. The current research was supported by Ministry of Education of Singapore grant 2021MOE-IF-024.